\documentclass[reprint,
nofootinbib,
 amsmath,amssymb,
 aps,
pra,
]{revtex4-2}
\usepackage{comment}
\usepackage{amsmath}
\usepackage{graphicx}
\usepackage{float}
\usepackage{hyperref} 
\usepackage{arydshln}
\hypersetup{colorlinks=true,urlcolor=blue,citecolor=blue,linkcolor=blue}

\begin{document}

\title{Student use of a quantum simulation and visualization tool}
\author{Shaeema Zaman Ahmed$^1$}
\thanks{These two authors contributed equally.}
\author{Carrie A. Weidner$^{1, 2*}$}
\email{c.weidner@bristol.ac.uk}
\author{Jesper H. M. Jensen$^1$}
\author{Jacob F. Sherson$^1$}
\author{H. J. Lewandowski$^{3, 4}$}

\affiliation{$^1$Department of Physics and Astronomy, Aarhus University, 8000 Aarhus C, Denmark}
\affiliation{$^2$Quantum Engineering Technology Laboratories, H. H. Wills Physics Laboratory and Department of Electrical and Electronic Engineering, University of Bristol, Bristol BS8 1FD, United Kingdom}
\affiliation{$^3$Department of Physics, University of Colorado, Boulder, CO 80309, USA}
\affiliation{$^4$JILA, National Institute of Standards and Technology and University of Colorado, Boulder, CO, 80309, USA}

\begin{abstract}
   Knowledge of quantum mechanical systems is becoming more important for many science and engineering students who are looking to join the emerging quantum workforce. To better prepare a wide range of students for these careers, we must seek to develop new tools to enhance our education in quantum topics. We present initial studies on the use of one of these such tools, Quantum Composer, a 1D quantum simulation and visualization tool developed for education and research purposes. In particular, we conducted five think-aloud interviews with students who worked through an exercise using Quantum Composer that focused on the statics and dynamics of quantum states in a single harmonic well system. Our results show that Quantum Composer helps students to obtain the correct answers to the questions posed, but additional support is needed to facilitate the development of student reasoning behind these answers. We also show that students are able to focus only on the relevant features of Quantum Composer to achieve the task.  
\end{abstract}

\maketitle

\section{Introduction}
\label{sec:Intro}

Due to the recent attention on quantum technologies and their economic impact as a result of the so-called second quantum revolution~\cite{Deutsch}, efforts are increasingly focused on educating the next generation of individuals who will make up the backbone of the quantum workforce. As such, current research is increasingly focusing on how higher education institutions can provide the preparation required for people to be successful in this emerging industry and to bring quantum technologies out of academic labs and into society~\cite{Heather_2020, Asfaw}. Additionally, the quantum workforce must draw from a wide range of backgrounds, requiring educators to build quantum literacy among individuals that have less mathematical background than the typical physics graduate.

To be able to meet this demand for a quantum literate workforce, the community needs to develop new quantum curricula at all levels \cite{Stadermann, Heather_2020, QSmartWorkforce}, effective methods of teaching quantum mechanics to address common student difficulties in learning the relevant material~\cite{Singh_2015, vanJoolingen_2017}, and tools that allow students to visualize and explore quantum systems (e.g., those described in Refs.~\cite{QUILTS, QUVIS, PhET_QM}). If implemented effectively, these efforts will allow students to gain the skills necessary to contribute to the development and deployment of quantum technologies.

Here, we present a study of a new quantum software tool known as Quantum Composer (or simply Composer)~\cite{Composer_Cookbook}. Composer is an interactive and flexible flow-based programming tool designed to allow users to simulate and visualize the static and dynamic properties of one-dimensional quantum systems, including systems that are not readily analytically tractable. This study, which is an extension of preliminary work~\cite{Carrie_2020}, was designed to identify how students used Composer to explore questions about 1D quantum mechanical systems.  Previous work suggested that students used the visualizations contained within Composer to develop their conceptual understanding of a problem. The work presented here aims to probe the use of Composer more deeply.

In this study, we conducted think-aloud interviews with five students as they worked through an exercise designed to guide them through the statics and dynamics of quantum states in a quantum harmonic well system, and here we present the results from the single-well exploration. 
Our study sought to answer the following research questions: 
\begin{itemize}
    \item RQ1: Did Composer help students determine correct answers and reasoning of problems in simple 1D quantum systems?
    \item RQ2: How did students interact with Composer to complete the tasks?  
\end{itemize}

This study demonstrates the productive use of a new quantum visualization and simulation tool, while presenting the advantages and limitations of the tool with regards to students' abilities to provide correct answers and reasoning while exploring 1D quantum mechanical systems. 
 
\section{Background}
\label{sec:background}
 
\subsection{Student learning of quantum mechanics concepts}

Given the non-intuitive nature of quantum mechanics that arises due to concepts that are absent in classical physics (e.g., wave-particle duality, the quantization of energy levels, tunnelling), physics education researchers have studied extensively where common gaps in knowledge or misconceptions (i.e., interpretations inconsistent with common undergraduate-level instruction in quantum mechanics) arise. In order to develop an exercise that adequately tested the use of Composer for its capabilities, its limitations, and student learning, we drew from this literature and selected a subset of topics that our study could address. This is briefly summarized below, with an emphasis placed on the elements that featured components relevant to our work, that is, the statics and dynamics of stationary and superposition bound states (i.e., not scattering states).

Early efforts to categorize student misconceptions in quantum mechanics were done by Styer~\cite{Styer_1996} and Singh~\cite{Singh_2001}. These studies focused on how quantum states are represented, evolve in time, and are measured. More comprehensive reviews of these student difficulties were recently compiled by Singh and Marshman~\cite{Singh_2015} and Krijtenburg-Lewerissa \emph{et al.}~\cite{vanJoolingen_2017}. In particular, both reviews discussed student difficulties with differentiating between a quantum state's energy, wavefunction, and probability density. Singh's review also found students had difficulties with answering questions about the time-dependence of systems; this result was confirmed by an earlier study by Cataloglu and Robinett~\cite{QMVI} that focused on testing conceptual understanding and ability to work with visual representations in quantum mechanics; this work was later expanded on by Chhabra and Das~\cite{Chhabra_2016, Chhabra_2018}. Singh found that this difficulty persists even beyond the introductory level and into graduate courses~\cite{Singh_2008}. This study and others by Zhu and Singh~\cite{Zhu_2009, Singh_2012} also suggest that students face challenges with sketching the shape of the wavefunction. In later work, Emigh \emph{et al.} indicated that students fail to see the difference between the time evolution of wavefunctions versus probability densities, and they struggle to interpret a wavefunction's time-dependent phase factor(s)~\cite{Emigh_2015}. Student difficulties in interpreting complex exponentials were also identified by Wan \emph{et al.}~\cite{Wan_2016}. Similarly, Emigh \emph{et al.} noted that students tended to misinterpret the physical meaning of the real and imaginary components of the wavefunction. This is supported by other work done by McKagan \emph{et al.}~\cite{Wieman_tunneling_2008} and Passante and Kohnle~\cite{Gina_2019}, and both of these studies were framed in the context of visualization tools and their potential to assist student learning. In the Supplementary Material (SM), we provide a detailed overview of quantum visualization and simulation tools other than Composer, which we describe next.
 
\subsection{Quantum Composer}

In this study, we investigate the student use of a new quantum simulation tool: Quantum Composer. Composer was developed at Aarhus University (where this study was performed) and is available for free download for all major desktop computer platforms~\cite{Quatomic}. Rather than focusing on individual, curated simulations like the three projects mentioned in the SM, Composer is a flexible and interactive tool that enables educators and students to build and simulate one-dimensional quantum systems through a `drag-and-drop' visual programming and execution interface. Composer is described in detail in Ref.~\cite{Composer_Cookbook} and relevant elements are discussed briefly here.

\begin{figure*}
  \includegraphics[width=\textwidth]{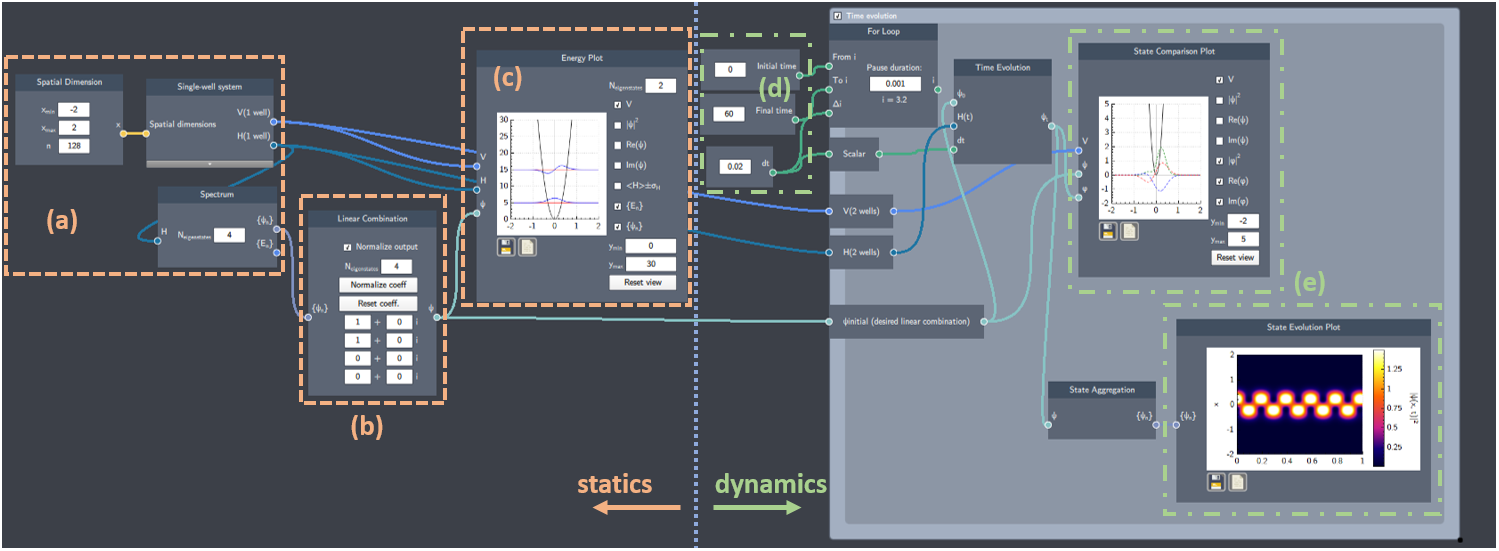}
  \caption{(Color online) A screenshot of the \emph{flowscene} for the single-well potential used by the interview participants. The blue dotted lines separate the time-independent parts of the simulation from the time-dependent dynamics that take place in the \emph{Time evolution} loop. All parameters shown in white boxes could be changed by the students. Students could observe the behavior of the system under study during the (a) the time-independent parts of the scenario (orange, dashed box within ``statics'') and (b) the time-dependent parts of the scenario (green, dashed box within ``dynamics''). The plots in (b) updated continuously if the student initiated time evolution by pressing the play button (not shown).}
  \label{fig:scenario_2}
\end{figure*}

The interface consists of a simulation environment where elements are dragged-and-dropped and connected together sequentially, as shown in Fig.~\ref{fig:scenario_2}. These elements are referred to as \emph{nodes} with interactive capabilities, e.g., entering scalars to create an arbitrary linear combination or defining a potential function. Composer also consists of a collection of visualization \emph{nodes}, such as the \emph{State Comparison Plot} that displays single eigenstates and superposition states, both in static and time-dependent configurations. The plots have checkboxes that enable the user to select which information they would like to see on the plot. For instance, one can visualize the real part, imaginary part, and the probability density of any wavefunction either simultaneously or one at a time by selecting the relevant checkbox. Composer is also capable of time-evolving a state given an initial state and potential. Once a simulation is built, it can be saved and loaded as a \emph{flowscene}, which can be used by students for guided exploration.

The purpose of this investigation is to explore the use of Composer as a tool for quantum visualization and simulation in an educational setting. Here, we extend previous developments on the use and impact of visualization and simulation tools in physics education research.
 
In this study, students answered questions related to the quantum harmonic oscillator based on an exercise (cf. Sec.~\ref{sec:Exercise}) given to them with an accompanying \emph{flowscene}. Their actions and thought processes during the exercise were captured through think-aloud interviews, and these data were analyzed to investigate the use and impact of Composer on student knowledge of quantum mechanics.

\section{Methods}
\label{sec:methods}

\subsection{Research context}\label{sec:research_context}

This study took place at Aarhus University, and the five students recruited for the study were enrolled in a second-year undergraduate quantum mechanics for nanoscientists course in Fall 2019. Students provided their informed consent prior to the study, and the consent form and study were approved by Aarhus University and is in compliance with the European Union General Data Protection Regulation. Course details and student exposure to Composer within the course can be found in the Supplementary Material, as can details on the interviews, which were done in the think-aloud style~\cite{TAP}.

\subsection{Interview protocol and exercise }
\label{sec:Exercise}

\begin{figure}
  \includegraphics[width=\columnwidth]{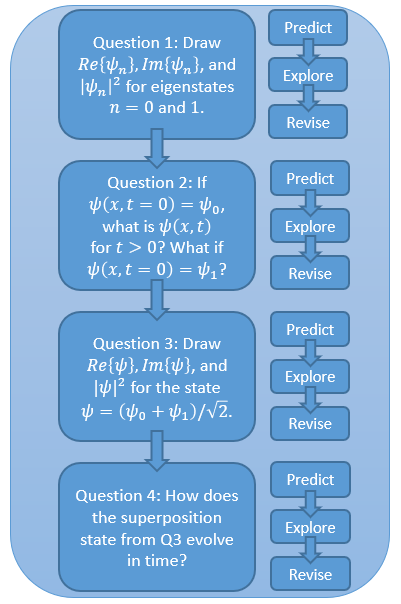}
  \caption{A simplified flowchart of the single-well exercise with the questions numbered as discussed in the Results section and the \emph{Predict}, \emph{Explore}, \emph{Revise} phases associated with each question. The exercise asked students to consider statics, then dynamics, of single eigenstates before moving on to superposition states.}
  \label{fig:flowchart}
\end{figure}

The exercise that the students worked through in the think-aloud interview covered statics and dynamics in two separate scenarios using single- and double-harmonic well potentials (see Supplemental Material for the full exercise prompt). In this study, we consider only the single-well part of the exercise, as shown in Fig.~\ref{fig:flowchart}.

For the interviews, we provided a previously constructed \emph{flowscene} for the participants to work with, as shown in Fig.~\ref{fig:scenario_2}. 
Additionally, we used the \emph{Scope} feature within Composer to hide unnecessary details of the simulation~\cite{Composer_Cookbook}, in order to encourage students to focus on the important aspects of the simulation and avoid unnecessary cognitive overhead that would hinder their exploration~\cite{Adams_2009}.

The exercise was designed to use a \emph{Predict, Explore, Revise} framework. That is, before using Composer, students were asked to \emph{Predict} (using either words, equations, or sketches) specific aspects of the system under study. After making their predictions, the students were then prompted to \emph{Explore} the same system in Composer and finally, they were asked to \emph{Revise} their answer and reasoning if needed. 

Students began the exercise by considering the eigenstates of the harmonic well. The exercise asked them to sketch the real part, imaginary part, and probability density of the ground state, $\psi_0(x)$, and first excited state, $\psi_1(x)$. Students were provided with a piece of paper with plots showing the harmonic oscillator potential on which they could sketch each state. After their sketches were completed, they then used Composer to produce plots of the same states and were prompted to compare the results they obtained in Composer with what they drew and revise any inconsistencies between their drawings and the Composer plots.

The exercise then prompted students to consider the time-dependence of the system. The exercise sheet reminded them that the $n$-th eigenstate (with energy $E_n$) evolves in time as
\begin{equation}
    \label{eq:time_ev}
    \psi_n(x,t) = \psi_n(x)e^{-iE_n t/\hbar}.
\end{equation}
The students were then asked to \emph{Predict} how the real part, imaginary part, and probability density of the ground and first excited states evolved in time, after which they went through the same \emph{Explore} and \emph{Revise} phases.

After considering single eigenstates, students then considered the superposition state
\begin{equation}
    \label{eq:superposition}
    \psi_\mathrm{lin}(x,t) = \sqrt{\frac{1}{2}}\big(\psi_0(x)e^{-iE_0 t/\hbar} + \psi_1(x)e^{-iE_1 t/\hbar}\big)
\end{equation}
 with the initial state at $t = 0$ given by
\begin{equation}
    \label{eq:superposition_init}
    \psi_\mathrm{lin}(x) = \sqrt{\frac{1}{2}}\big (\psi_0(x) + \psi_1(x) \big).
\end{equation}
First, students considered the static case, and as before, they were asked to \emph{Predict} the real part, imaginary part, and probability density of $\psi_\mathrm{initial}(x)$. However, they were also asked to \emph{Predict} whether or not the probability density of the state would evolve in time and why. Again, students were asked to \emph{Explore} and \emph{Revise} their predictions using Composer. This marked the end of the single-well part of the exercise.

\subsection{\label{sec:coding}Coding scheme}

After all interviews were completed, the audio was transcribed by humans via an online transcription service~\cite{rev}. The video, audio, and transcripts were all used during the coding process. In this context, coding refers to categorizing and labeling the students' vocalizations and on-screen actions. Extensive details on the coding scheme and the codebook used can be found in the Supplemental Material, and we include relevant aspects here.

In addition to coding the interviews, we documented the accuracy of students' responses to the questions in the \emph{Predict} phase to capture whether a prediction and the reasoning surrounding it were correct (including partially correct),  completely incorrect, or not present. We included partially correct answers among the correct answers as there was a rather large spectrum of ``correctness'' and it was difficult to place an answer on that spectrum, e.g., due to students' difficulties with the language of quantum mechanics.

\section{Results}\label{sec:results}

We address our research questions (outlined in Sec.~\ref{sec:Intro}) in separate subsections. 
All exercise questions referred to in this section are labelled and briefly described in the flowchart in Fig.~\ref{fig:flowchart}.

\subsection{Impact of Composer on student answers and reasoning} \label{sec:progression}

\begin{figure}
  \includegraphics[width=\columnwidth]{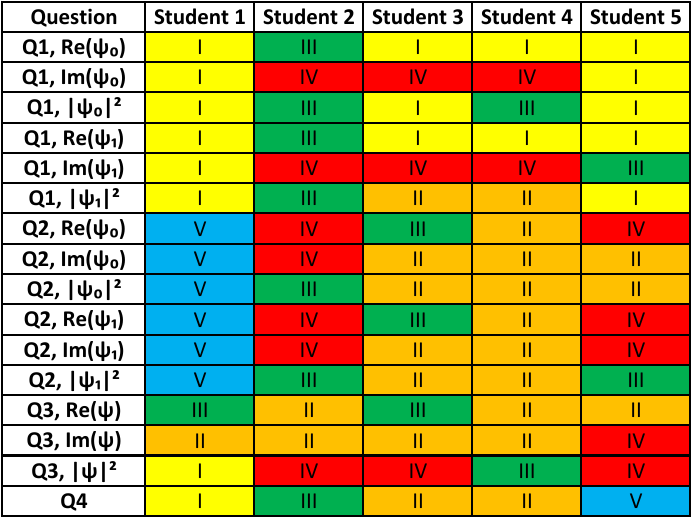}
  \caption{Categorization of individual student responses for each question (cf. Fig.~\ref{fig:flowchart}), color- and Roman-numeral-coded by outcome: (I, yellow) corresponds to the case where the student gave the correct answer in the \emph{Predict} phase but provided no reasoning at any stage; (II, orange) corresponds to the case where students provided a correct answer with correct reasoning in the \emph{Predict} phase; (III, green) denotes where students provided correct reasoning after (but not before) using Composer in the \emph{Explore/Revise} phase; (IV, red) marks cases where (a) the initial answer was incorrect or not provided and no reasoning was provided upon revision or (b) no answer was provided in the revision stage; and (V, blue) shows where students used Composer incorrectly at any stage of the process.}
  \label{fig:trajectories}
\end{figure}

Here, we address RQ1: does Composer help students determine correct answers and associated reasoning of problems in simple 1D quantum systems? To this end, we present in Fig.~\ref{fig:trajectories} categorization of students' responses to each question asked in the exercise. Additional details for each question are available in the SM.

As described in Sec.~\ref{sec:coding}, we categorized students' responses to the exercise questions as `correct' if they were entirely or partially correct, `incorrect' if the responses were completely incorrect, or `not present' if no response was provided. This categorization, when applied during the \emph{Predict} phase indicates the measure of students' knowledge about the concept before using Composer, and when applied in the \emph{Explore/Revise} phase indicates if Composer facilitated any improvement to their responses.

Importantly, students were not explicitly asked to revise initially correct answers that were provided without reasoning, and most students would provide correct answers without reasoning for some questions in the exercise. We note, however, that unless an answer was not provided in the \emph{Explore/Revise} phase (which only occurred once), all students were able to provide a correct answer upon using Composer (provided they were using it correctly), indicating that students can, at the very least, use Composer to obtain the correct answer. Regarding the correct use of Composer: Student 1 used Composer incorrectly during the first question that discussed time-evolution (Q2) due to the fact that they were not aware of the \emph{Play} button that would show wavefunction dynamics. Likewise, Student 5 used the static (not dynamic) plot to answer Q4.

When examining at these data question-by-question, we see that difficulty with the imaginary part of Q1 was common.
For example, Student 4 stated
\begin{quote}
    \emph{I can't actually remember...about the imaginary part. I don't remember if there is a imaginary part, so I'll just skip that and maybe think about it later.}
\end{quote}
The student was, in the \emph{Explore/Revise} phase, able to identify that the imaginary part of the wavefunction was zero, and this knowledge was correctly applied in later \emph{Predict} parts of the exercise. Interestingly, after looking at Eq. (2) while in the \emph{Predict} phase for Q2, the student was able to apply correct mathematical reasoning to answer why the superposition state would have zero imaginary part in Q3. All but one of the other students were able to do the same, e.g., where Student 1 states
\begin{quote}
    \emph{So, as t increases, the exponential factor gets smaller. That's imaginary. Okay. I've only got the time and the exponential factor. We're introducing an imaginary part, so the imaginary part won't be zero anymore, I guess, at least.}
\end{quote}

The next question in the exercise involved time evolution. Unlike in the previous question, most students provided correct answers and reasoning for the imaginary part of the wavefunction in Q2, but we note that they were given the mathematical representation of the wavefunction in the exercise. An example of such a response is where Student 1 states
\begin{quote}
    \emph{So, as t increases, the exponential factor gets smaller. That's imaginary. Okay. I've only got the time and the exponential factor. We're introducing an imaginary part, so the imaginary part won't be zero anymore, I guess, at least.}
\end{quote}

On the other hand, students tended to make incorrect predictions for the time evolution of the real part of an eigenstate, even if they were all able to predict the $t = 0$ wavefunction (Student 2 had a correct prediction without reasoning, but then provided correct reasoning upon revision). Many students predicted the real part to be static in time. Student 3 was one of these, but after using Composer, the student said
\begin{quote}
\emph{Of course it's changing in time as well, the real part...Because now we have a time dependency...for the real part, I actually didn't even think about the time dependency of the imaginary part working here on the real part. So it was actually nice seeing that. It had an influence because sometimes when I'm just looking at the equations it doesn't give you the interpretation of the wave functions.}
\end{quote}
Here, the student realized that the real part is time-dependent after looking at the dynamic \emph{State Comparison Plot} in Composer. Additionally, the student mentioned that it was not obvious earlier that the real part would evolve in time, even though one may know the equations. The student was able to connect the visualizations of the wavefunctions shown in Composer with the underlying mathematics describing the physics of the system. 

Similarly, Student 2 initially did not predict that the probability density of the ground state would be stationary in time. Using Composer helped them to recall what they had learned in class, as shown below:
\begin{quote}
\emph{Oh, I need the probability density...and here we go. Let's see. There's something wrong...why isn't that moving more?...Oh, okay. So this part, I thought earlier would be moving. [It] doesn't move because it's the probability. But it makes sense, actually, from the theory that I have learned.}  
\end{quote}
Later in the exercise, when asked to predict the dynamics of the first excited state, Student 2 indicates a change in thinking, stating 
\begin{quote}
    \emph{Now I have learned that the probability density would also look the same [as time evolves].}
\end{quote} 
Thus, the student has recalled additional knowledge in the earlier part of the exercise and applied this to make a correct prediction in another part of the exercise.

When considering a static superposition state, many students were able to determine the real and imaginary parts of the state, some with the help of Composer. However, the probability density proved more challenging to reason about. One notable exception is given by Student 4, who states, upon using Composer:
\begin{quote}
    \emph{I can see that I was quite wrong, and it also makes sense because I can't just add the probability like before because that's not how it works. For the wavefunctions, I have to be able to draw it properly, and I can see it doesn't quite look like that. And then, I just I have to use this to make the norm square and to get the right answer, so I was quite wrong...I have learned that I just can't look at the probability density from the two eigenstates and then add it together.}
\end{quote}
The student has used the visualizations from Composer to correctly revise their reasoning about the correct answer in that they remember that they must take the modulus squared of the superposition state instead of adding the probability densities of the individual eigenstates. Thus, Composer facilitated their reasoning by making them confront the conflicting answers.

Holistically, we see that while Composer can facilitate the development of correct reasoning, this is not universal, as indicated by the roughly equal number of items in Fig.~\ref{fig:trajectories} corresponding to cases III (correct reasoning provided after using Composer) and IV (no reasoning provided upon revision or no answer provided upon revision).

\subsection{Composer features}
\label{sec:features}

In this section, we address RQ2: How did students interact with Composer to complete the tasks?  We do this to understand to what extent students are exploring all of the available Composer features or focusing only on the relevant aspects of the \emph{flowscene}, and whether or not they focus only on one or two relevant features. 
 
\begin{figure*}[htbp]
    \centering
    \includegraphics[width=\textwidth]{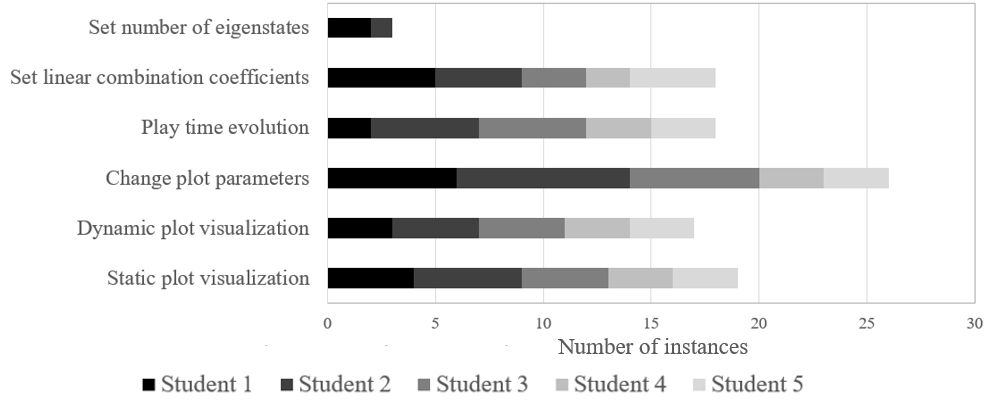}
    \caption{Bar chart showing the number of separate instances coded for each student's use of various Composer \emph{Interactions} and \emph{Visualizations} during the \emph{Explore/Revise} phase of the single-well system exercise. The different colored bars in the charts show how often each student performed each \emph{Interaction} or made use of each \emph{Visualization}, and the stacked sum of all bars shows the total number summed across all students.}
    \label{fig:Feature_instances}
\end{figure*}

We quantified the use of Composer features by counting the number of separate instances coded for the different features. In this section, we consider six subcodes under the \emph{Exploration} code (denoted by * in the codebook, see the SM), as these are the features used during the exercise. We present the cumulative number of instances that were coded in all five interviews in Fig. \ref{fig:Feature_instances}.

The graph indicates that students engage regularly with all features with the exception of \emph{Set number of eigenstates}, which did not need to be changed to complete the exercise. We see that students work more extensively with the relevant features, including those related to visualization of the system, using each about the same number of times.

\section{Discussion and Implications for Instruction}
\label{sec:discussion}

These results show that Composer allows students to visualize the correct behavior of a given system and facilitates their ability to make connections between what they learned in their coursework with the visualizations obtained from Composer. Additionally, by using Composer, students were able to better connect the underlying mathematics with the visualizations they were seeing.  As they went through the exercise, almost all students were able to express the correct answers after using Composer, and a number of students indicated that the tool is a useful way to check their answers. Below, we discuss in more detail the results related to our research questions. Limitations of the study can be found in the SM.

First, we discuss our results relevant to RQ1: does Composer help students determine correct answers and associated reasoning of problems in simple 1D quantum systems? Perhaps most interestingly, our results show that students' use of Composer and the mathematical forms presented in the exercise (cf. Eq.~\ref{eq:superposition}) helps them to reason about the imaginary part of wavefunctions as they move through the exercise. This is particularly compelling, as it is common for students to struggle with visualizing the imaginary part of the wavefunction~\cite{Emigh_2015, Wan_2016}, and this shows that, when used alongside the mathematical expressions, students can use Composer to reason about this concept and correctly apply their knowledge as they move through an exercise. This is supported by the following student quote:
\begin{quote}
    \emph{And the imaginary part, yeah, so it also is constant at time equals zero. Okay. I'll just revise. So I think it's getting better, because I now have the knowledge from Composer, how they actually evolve, now I was able to answer more correctly than before. So I think it's working as it should.}
\end{quote} 
Therefore, we believe that Composer could be particularly beneficial in addressing questions concerning the time evolution of states, as it allows students to visualize the time evolution for the real part, imaginary part, and the probability density; this can assist with the difficulties identified in Refs.~\cite{Singh_2015, vanJoolingen_2017}. Moreover, features like the \emph{State Comparison Plot} (coded as \emph{Dynamic plot}) help students compare the behavior of a quantum state in the static and dynamic cases. This assists students in determining the differences between the two cases, although Composer alone is not necessarily well-suited to helping students to express their reasoning behind their answers, a finding echoed in other work on engaging students in electronic learning tutorials~\cite{Singh_2017}. Part of this is due to how our exercise was worded in that students who expressed the correct answer initially were not asked to reason about the problem upon revision, and we found that many students would simply provide correct answers without reasoning, particularly in the initial phases of the exercise.

Thus, Composer is likely, as with most other quantum visualization and simulation tools, best used in conjunction with a textbook, exercise, lecture, or other tool that focuses on elucidating the reasoning behind the physical behavior of the system under study, as is done in Refs.~\cite{Rainer_2002, Zollman, Physlet, PhET_QM}. For example, student reasoning may be better developed in an exercise where students work out a problem using mathematical representations and check their answers in Composer before being asked to reason about the problem. This can help them to build insight into problems where they connect what they see in Composer with what they have worked out mathematically. Additionally, these exercises should help students develop better reasoning through guiding questions that connect the math to the visualizations~\cite{Wilcox_2013}.

We note, however, that when students use Composer, they can get the wrong answer, which, in the worst case, can facilitate the development of incorrect reasoning. This indicates the need for comprehensive onboarding during Composer use in the classroom, and instructors should look out for students who are using the tool incorrectly.

Now, we briefly discuss the results related to RQ2: How did students engage with Composer to complete the tasks? Our finding show that students generally focus only on relevant aspects of the Composer \emph{flowscene}, suggesting that Composer is laid out in such a way to support student use of the tool. In addition, students focus largely on the visualization aspects provided in Composer (i.e., the last four lines in Fig.~\ref{fig:Feature_instances}), which suggests that they are using the tool as intended, that is, to visualize the system at-hand. Thus, Composer as-is is sufficient to aid students in their exploration of 1D quantum physics, although this does not rule out the possibility of future improvements that can be made to the interface.

\section{Conclusion and outlook}
\label{sec:conc}

The aim of this study was to probe whether Quantum Composer can be used as a tool to increase students' knowledge of quantum mechanics in a single harmonic well system. The exercise covered questions related to the static and dynamic case of the real part, imaginary part, and probability density of single eigenstates and superposition states. The study used a think-aloud protocol with five students, where we found that all students could determine or revise answers to problems in quantum mechanics by changing parameters and visualizing the system in Composer. Overall, we were able to demonstrate the constructive use of a new quantum visualization and simulation tool and show how student responses to questions about quantum systems was aided by exploration in Composer, in particular, with regards to their ability to provide the correct answers to questions posed.

This work bolsters previous studies on the impact of quantum mechanics visualization tools. First and foremost, the use of tools like Composer can facilitate students' ability to articulate the correct answer, but more effort is required to ensure that students are properly reasoning about the behavior of a given quantum system. Thus, we recommend that the use of such tools be coupled directly to other course materials and environments (e.g., guided tutorials, homework exercises) that help students connect their answers to the underlying reasoning.

As this was just our first exploratory study of the benefits and limitations of using Composer to aid in student learning of 1D quantum mechanical systems, there are many opportunities for more in-depth research studies. Future work will include a larger number of student participants with different demographics and educational backgrounds. Additionally, we could examine clickstream-type data to better understand how exactly students interact with all of Composer's features.  We could also create new types of exercises that help develop and elicit student reasoning around these topics. Given Composer's unique flexibility, we could also study how students and educators create and/or modify \emph{flowscenes}, as they illustrate or explore different concepts. Ultimately, we hope to be able to not just refine Quantum Composer so that it is more useful for students, but we also hope to understand the best way to incorporate Composer in quantum mechanics courses for all science and engineering majors, including those focused on a career in quantum information science and technology. 

\acknowledgments

{The authors would like to thank M. Murdrich for his support regarding the use of Composer in the course. We acknowledge funding from the European  Union's Horizon 2020 research and innovation programme under the Marie Sk\l{}odowska-Curie QuSCo  grant agreement No. 765267 and the ERC Proof-of-Concept grant PQTEI. Additional support was provided by the US National Science Foundation (PHY-1734006 and QLCI Award OMA-2016244).
}

%\newpage

%%%%%%%%%% Merge with supplemental materials %%%%%%%%%%
\pagebreak
\widetext
\begin{center}
\textbf{\large Student use of a quantum simulation and visualization tool, supplementary material}
\end{center}
%%%%%%%%%% Merge with supplemental materials %%%%%%%%%%
%%%%%%%%%% Prefix a "S" to all equations, figures, tables and reset the counter %%%%%%%%%%
\setcounter{equation}{0}
\setcounter{figure}{0}
\setcounter{table}{0}
\setcounter{page}{1}
\makeatletter
\renewcommand{\theequation}{S\arabic{equation}}
\renewcommand{\thefigure}{S\arabic{figure}}
\renewcommand{\bibnumfmt}[1]{[S#1]}
\renewcommand{\citenumfont}[1]{S#1}
%%%%%%%%%% Prefix a "S" to all equations, figures, tables and reset the counter %%%%%%%%%%

\section{Quantum visualization and simulation tools}

Studies have shown that visualization and simulation tools help facilitate student learning in quantum mechanics in a multitude of ways: by making connections with pre-existing knowledge~\cite{QMVI}, building mental models~\cite{QUVIS}, and developing their ability to work with visual representations of quantum physics concepts~\cite{Passante_2019, Kohnle_2010}. Such tools also present numerical calculations, allowing students to visualize complex phenomena and systems beyond what is analytically tractable, including multidimensional systems and time-dependent behavior~\cite{QMVI}.

The use of computer-based visualization and simulation tools in quantum mechanics education has its roots in a series of books published in 1995-6 by the Consortium for Upper Level Physics software, two of which focus on quantum mechanics~\cite{CULPS_1, CULPS_2}. Other such software-textbook combinations have been published~\cite{QMS,Thaller, Thaller2}. More recently, the Physlet applets were developed by Belloni \emph{et al.}~\cite{Physlet}, and similar tutorial-style units with integrated simulations have been developed~\cite{Zollman, Rainer_2002}. There also exist other online resources for interested students and instructors~\cite{Falstad, VQM, Quantum_online}. Additionally, one can simulate quantum phenomena through the use of text-based programming platforms like MATLAB, Python, and C++, and several quantum-specific libraries for these platforms are available, among them those described in Refs.~\cite{QuTiP, QuTiP_2, QEngine}.

However, the three main research-based quantum visualization tools that are being actively developed (besides Composer) are the PhET simulations developed at the University of Colorado Boulder~\cite{PhET_website}, the quantum interactive learning tutorials (QuILTs) developed at the University of Pittsburgh~\cite{QuILTs_website}, and the quantum mechanics visualization simulations (QuVIS) developed at St. Andrews University~\cite{QuVIS_website}. Each of these tools is modular in the sense that they present individual simulation modules typically addressing a carefully chosen subset of concepts, unlike Composer in which one can build a large variety of customizable scenarios from scratch. They have all been demonstrated to enhance learning of quantum mechanics at the undergraduate level and are freely available online.

The first of these tools, the PhET interactive simulations, is a suite of over 150 simulations covering multiple STEM fields, of which 21 are currently categorized under \emph{quantum phenomena}. These cover topics like band structure, the structure of the hydrogen atom, and the Stern-Gerlach experiment. Extensive research has gone into the development and use of these simulations, both in general~\cite{PhET_general_1, PhET_general_2, PhET_general_3, PhET_general_4} and specifically for quantum mechanics~\cite{PhET_QM, Wieman_tunneling_2008}.

The second tool, the QuILTs, has been described in detail in Ref.~\cite{QUILTS}. The QuILTs are a set of 14 Java applets designed to help students use computer-based visualization tools coupled with curriculum elements like tutorials, tests, and homework assignments to tackle the challenging topics identified through research~\cite{Singh_2001, Singh_2015}. Student engagement with these tutorials is covered in detail in DeVore \emph{et al.}~\cite{Singh_engagement_2017}. They suggest that such tools can be powerful in focused, one-on-one interview sessions, but, in some cases, they may be used more superficially (i.e., less effectively) in unsupervised self-study sessions. QuILT topics are applicable for a broad range of educational levels from the introductory to the advanced undergraduate, including expectation values in quantum mechanics~\cite{Singh_2017}, the double-slit experiment~\cite{Singh_double_slit_2017}, and degenerate perturbation theory~\cite{Singh_DPT_2019}.

Finally, the QuVIS project has developed numerous research-based simulation applets, typically accompanied by an activity sheet, for students at the high school to advanced undergraduate level~\cite{Kohnle_2010, QUVIS}. Topics covered by the QuVIS project include the time-dependence of states~\cite{Gina_2019}, two-level systems~\cite{Kohnle_2015}, perturbation theory~\cite{Kohnle_2017}, and single-photon superposition states~\cite{Kohnle_PERC}, among many others.

\section{Course details}

In Fall 2019, 31 students were enrolled in the quantum mechanics for nanoscientists course. Note that Aarhus University is the second-largest university in Denmark, with approximately 38,000 students enrolled in bachelor's, master's, and PhD programs. The student population is predominantly white and native to Denmark. The quantum course was taught in English, and the study was in English, but all of the students who volunteered for our study spoke English as a second language. Students were recruited in-class and via messages posted on the course's online message board, and volunteers were asked to give two hours of their time. Participants were compensated with a gift card worth 500 Danish kroner. The think-aloud interviews took place in the weeks following the final exam. All interviews were conducted by S.Z.A, and students were given one hour to complete the full exercise (single- and double-well).

After the interviews, the students were asked some basic questions on their background, thoughts about Composer, and gender information. Participants were told that a response to the gender question was optional, but all students responded. Three men and two women participated in the interviews.

In addition to instruction in the quantum mechanics for nanoscientists course, four of the five students reported also having had some introductory quantum mechanics in a physical chemistry course taken concurrently with the quantum course. In addition to courses covering quantum topics, nanoscience students typically take courses covering mechanics, thermodynamics, calculus, electrodynamics, waves, and optics prior to their quantum course.

The quantum course was taught using Griffiths' textbook~\cite{Griffiths}, covered the quantum harmonic oscillator extensively, and included seven homework assignments based on the book Several of the problems required students to use programming tools like MATLAB. Three times throughout the course, the students were given additional activities built around Composer \emph{flowscenes} (see Fig.1 in main text); these activities were designed by C.A.W., S.Z.A., and J.H.M.J. None of the activities were graded, nor mandatory. The Composer activities were designed so that students did not have to build anything from scratch in Composer (although they were shown how this could be done). The first set of activities explored the infinite and finite square well, the second set investigated superposition and expectation values, and the third discussed time-independent perturbation theory. None of the Composer activities explored time dependence, but time-dependence was covered in the course. Before the first Composer activity was given in the course, students were given a 15-minute presentation on how to use Composer. For all Composer activities, one of the authors was present for approximately 30 minutes to answer questions. Aside from the Composer elements in the quantum course, students did not report significant simulation-based activities in past courses.

\section{Think-aloud interviews}

In order to examine whether and how students provide answers and reasoning to problems in the quantum mechanics exercise for this study, a think-aloud protocol was used. The think-aloud protocol involves speaking aloud the thoughts that come to the interviewee's mind while performing the task at hand~\cite{TAP}. This methodology is used to provide rich verbal data describing how one tackles a problem-solving task, which provides inferences about the the steps taken to solve the problem~\cite{TA}. Therefore, for the purpose of our study, this methodology was considered to be appropriate. The potential of this methodology has been explored for education action-research~\cite{TAstudy2} and has 
been applied to different educational research studies as well, both inside~\cite{Heckler, Passante_2019, Heather_2019, troubleshooting} and outside of physics~\cite{TAstudy}. 

In the context of this study, the think-aloud protocol was used to collect data on how students work through the different components of the exercise. Interviews were conducted with five students, and during each interview, the student volunteer verbalized their thoughts aloud while working through the exercise. During this process, the students' voices and computer screens were recorded.  Additionally, the students' sketches and notes were also documented during the interview. We recorded each student working through the exercise for 55-65 minutes, followed by answers to background questions about Composer and demographics for approximately 5-10 minutes. Before the start of the interview, the students provided their written consent to this study and were briefed about the think-aloud format. Throughout the interview, the interviewer intervened in only two cases: either to remind the student to talk aloud if there were long periods of silence or if the student encountered technical problems with using Composer, e.g., not being able to find the \emph{Play} button or zooming in and zooming out on the screen.

 %The think-aloud protocol was chosen because it would elicit student reasoning about the problems at hand as they worked through the different phases of the exercise. Thus, even if students were not asked after every question to explicitly vocalize their reasoning, we still hoped to capture their thinking because of the chosen methodology.

\section{Coding scheme}

Sections of the interview were examined (including audio, video, and transcripts), and the audio-visual data was coded using the NVivo software package in one-minute increments. The audio transcripts and student sketches were used as supporting information during the coding process. The process of creating and refining the codebook was done collaboratively by C.A.W. and S.Z.A.,  with input and consistency checks from H.J.L. Similar video coding methods have been used in other work in physics education research~\cite{PhET_general_2, troubleshooting, Heather_2019}, and our codebook development and coding process are modeled on common and accepted methods for coding qualitative data in physics education research~\cite{Otero}.

The final codebook is shown in Tab.~\ref{tab:codebook}. To begin, the codebook was divided into three parts, corresponding to the \emph{Predict}, \emph{Explore}, and \emph{Revise} phases of the exercise. We began with an \emph{a priori} codebook of main codes (in bold in Tab.~\ref{tab:codebook}) that were driven by the exercise protocol and the specific Composer-related behaviors that we wanted to capture (i.e., student use and understanding of the tool). From the data obtained during iterative rounds of interview coding, we determined the emergent codes and subcodes (unbolded in the table) that best captured the students' behavior during the exercise and informed our research questions. Initially, students' behavior during the \emph{Explore} and \emph{Revise} phases was coded separately, but after the first coding iteration, these were combined as students would typically verbalize their revision in the \emph{Explore} phase. Given this, we describe the codes contained in Tab.~\ref{tab:codebook} in what follows. 

\begin{table}[ht!]
\caption{Coding scheme used in the study to code one-minute intervals of each think-aloud interview covering the \emph{Predict} and \emph{Explore-Revise} phases of the exercise. The codes denoted by an * are the relevant Composer features for the single-well system as discussed in Section IV.B of the main text. }
    \begin{tabular}{p{1cm} p{1cm}}
\hline
\textbf{Main code}& \textbf{Subcode}\\
\hline
\multicolumn{2}{c}{\emph{Predict}}\\
\hline
\textbf{Part of Exercise} & Single-well\\
 & Double-well\\
 & Single eigenstate\\
 & Superposition state\\
 & No time evolution\\
 & Time evolution\\
\textbf{Tool/Representation} & Math \\
 & Sketches\\
 \hline
 \multicolumn{2}{c}{\emph{Explore/Revise}}\\
 \hline
 \textbf{Part of Exercise} & Single-well\\
 & Double-well\\
 & Single eigenstate\\
 & Superposition state\\
 & No time evolution\\
 & Time evolution\\

\textbf{Tool/Representation} & Composer\\
 & Math \\
 & Sketches \\
\textbf{Exploration} & \\
Visualization
& Static plot*\\
& Dynamic plot*\\

Interaction & Change plot parameters* \\
 & Play time evolution*\\
 & Set linear combination*\\
 & Set number of eigenstates*\\

\textbf{Understanding of Composer} &\\
& Orientation\\
& Asking for help\\
& Receiving help\\
& Using Composer incorrectly\\
\hline
\hline
\end{tabular}\label{tab:codebook}
\end{table}

In our final codebook, the \emph{Part of Exercise} code category was used simply to label which part of the exercise the student was working with for future reference when doing data analysis. As students were working through the exercise, they used sketches, math, and Composer to explain their responses. Hence, the \emph{Tool/Representation} codes denoted the tool(s) and/or representation(s) being used as they were working through the exercise. When using Composer, students engaged with the tool and explored the physical system under study through visualizations (plots) and interactions like changing plot parameters or values in relevant fields. This exploration was coded under the \emph{Exploration} code category divided into the  \emph{Visualization} and \emph{Interactions} subcodes.

The \emph{Visualization} subcodes were used to identify when a student was using a visualization within Composer to answer a question, e.g., describing what was going on within a plot or watching the system evolve in time. In order for this code to be applied, we required that the interaction be obvious, e.g., the student was moving the mouse around the plot while discussing its contents or the student was describing the wavefunction's behavior as it evolved in time or the student changed parameters (like the spacing between the wells in the double-well system). The \emph{Interaction} code was applied when students were actively interacting with Composer elements and changing the parameters set in the program.

In situations where students were trying to understand the workings and features of Composer itself (e.g., how to play the simulation), this was coded under \emph{Understanding of Composer}, including situations where students were simply using Composer incorrectly, e.g., not pressing the \emph{Play} button to begin time evolution.

Once the final codebook was agreed upon after multiple iterations of individual-researcher coding, discrepancy reduction, and codebook modifications by C.A.W. and S.Z.A., all five interviews were coded  collaboratively coded by C.A.W. and S.Z.A., with input from H.J.L.

\section{Limitations}
\label{sec:Limitations}

One of the main limitations of this study is that the inferences drawn may not be generalizable with respect to claims about enhanced knowledge, as the study covered think-aloud interviews of only five students from one course at one university. This university was also the one where Composer was developed. We also have no information on how well the students who participated in the study performed in the course, so we do not know if our method of selection was biased towards students with high or low grades, although we did make it clear during recruitment that all students could participate regardless of their performance in the class. Additionally, English was not the first language of the interviewees, and this may have hindered the efficacy of the think-aloud protocol and our ability to discern their thinking. We also note that as this was the first such study using a brand new visualization tool, the data collected do not allow us to probe in-depth the exact nature of how students used Composer to facilitate their learning, but allow us to make only broad claims.

Finally, we found that students sometimes use Composer incorrectly and can thus misinterpret the resulting (erroneous) visualizations. This indicates the presence of a learning curve with Composer in that students must familiarize themselves with the tool before they can adequately use it to answer questions. For example, there were cases where students had trouble finding the play button to begin time evolution or using the linear combination node. In the former case, this could be due to the fact that none of the in-class Composer exercises explored time dependence. This informs future design changes in the introduction of students to Composer and to the interface itself; while this was not a goal of the current study, it is an outcome, and other tools have used similar studies to inform adjustments to their interfaces~\cite{PhET_QM}. Generally speaking, a comprehensive introduction is likely a prerequisite to let the user gain familiarity with Composer before diving in for exploration~\cite{Composer_Cookbook}.

That is, when developing such tools, building an intuitive interface can be challenging, and as a result, users should be onboarded to a given tool via a reference manual, an in-class demonstration of the tool, and/or a tutorial video. We discuss Composer onboarding more in Ref.~\cite{Composer_Cookbook}. Thus, we recommend using studies like those performed in Refs.~\cite{PhET_general_2, PhET_general_3} to inform interface design considerations.

\section{Exercise}

Note that the original exercise also included a double-well portion. We include this here for completeness, but this part of the exercise was not evaluated in the study. Furthermore, the \emph{Compare} section 
%directly after the static eigenstate problem. This part of the exercise asked students to discuss whether or not $\psi_0(x)$ and $\psi_1(x)$ were orthogonal, but in retrospect, this 
was determined to be confusing and difficult to determine in the \emph{flowscene} provided to the students. This type of exploration is not difficult to do in Composer, but the required nodes were not present in what the students used. Therefore, this part of the interview was not coded or considered in the analyses.

\subsection*{Introduction}

The purpose of this exercise is to understand how students use Composer to think about quantum mechanics. In this exercise, we will investigate what happens to the energy levels and states of the system as we move from a single oscillator well to two oscillator wells separated by some distance. The exercise has two parts - the single-well system and the double-well system which will be described below.

\subsection*{The single-well system}

As you saw in class, the Hamiltonian describing a harmonic oscillator is
\begin{equation}
H = T + V(x) = -\frac{\hbar^2}{2m}\frac{\partial^2\psi}{\partial x^2} + \frac{1}{2}m\omega^2x^2
\end{equation}
where $m$ is the mass of the particle and $\omega$ is the harmonic oscillator frequency.

A Composer file has been opened for you where you will investigate the energy levels and states of this single-well system.

\subsubsection*{1: States in the single-well system}

\begin{itemize}
\item{\bf{PREDICT} (without using Composer)}
    \begin{itemize}
        \item For the single-well system, draw (on the axes provided to you) the wavefunctions $\psi_n(x)$ that correspond to the ground ($n=0$) and first excited state ($n=1$). If you are not sure what to draw, take your best guess or leave it blank. For each case, draw the 
        \begin{itemize}
            \item The real part of the wavefunction.
            \item The imaginary part of the wavefunction.
            \item The probability density $|\psi_n(x)|^2$.
        \end{itemize}
    \end{itemize}

\item{\bf{EXPLORE} (using Composer)}
    \begin{itemize}
        \item Check your answer in Composer. Does this agree with what you drew?
    \end{itemize}
\item{\bf{REVISE}}
    \begin{itemize}
        \item If your prediction was incorrect, can you use the visualizations from Composer that you explored in the previous question to re-think your understanding?
    \end{itemize}
\item{\bf{COMPARE}}
    \begin{itemize}
        \item Now, using these representations, can you determine whether or not the ground state and first excited state are orthogonal? \item How do you know?
    \end{itemize}

\end{itemize}

\subsubsection*{2: Time evolution in the single-well system}

Now we will look at the time-dependence of the states of this system. Mathematically, the $n$th eigenstate $\psi_n(x)$ evolves as
\begin{equation}
    \psi(x,t) = \psi_n(x)e^{-iE_nt/\hbar}.
\end{equation}
where $E_n$ is the energy of the state, \textit{n}.

\begin{itemize}
\item{\bf{PREDICT} (without using Composer)}
    \begin{itemize}
        \item Using words, pictures, or equations, answer the following question: if you start in the \textbf{ground state} of the system, how do the following parts of the wavefunction change in time:
        \begin{itemize}
            \item The real part?
            \item The imaginary part?
            \item The probability density $|\psi_n(x)|^2$?
       \end{itemize} 
    \end{itemize}
\item{\bf{EXPLORE} (using Composer)}
    \begin{itemize}
        \item Check your answer in Composer. Does this agree with what you predicted?
    \end{itemize}
\item{\bf{REVISE}}
    \begin{itemize}
        \item If your prediction was incorrect, can you use the visualizations from Composer that you explored in the previous question to re-think your understanding?
    \end{itemize}
\end{itemize}  

Now we will repeat this exercise, but this time we will start in the \textbf{first excited state}.

\begin{itemize}
\item{\bf{PREDICT} (without using Composer)}
    \begin{itemize}
        \item Using words, pictures, or equations, answer the following question: if you start in the \textbf{first excited state} of the system, how do the following parts of the wavefunction change in time:
        \begin{itemize}
            \item The real part?
            \item The imaginary part?
            \item The probability density $|\psi_n(x)|^2$?
       \end{itemize} 
    \end{itemize}
\item{\bf{EXPLORE} (using Composer)}
    \begin{itemize}
        \item Check your answer in Composer. Does this agree with what you predicted?
    \end{itemize}
\item{\bf{REVISE}}
    \begin{itemize}
        \item If your prediction was incorrect, can you use the visualizations from Composer that you explored in the previous question to re-think your understanding?
    \end{itemize}
\end{itemize}

\subsubsection*{3 and 4: The superposition state}

For the next part of the exercise, let's set up a \textbf{linear superposition} of two states. Mathematically, this looks like
\begin{equation}
\psi_\mathrm{initial}(x) = {\sqrt\frac{1}{2}(c_\mathrm{0}\psi_0(x) + c_\mathrm{1}\psi_1(x))}.
\end{equation}
This state evolves in time as
\begin{equation}
\psi_\mathrm{initial}(x,t) = {\sqrt\frac{1}{2}(c_\mathrm{0}\psi_0(x)e^{-iE_0t/\hbar} + c_\mathrm{1}\psi_1(x)e^{-iE_1t/\hbar})}.
\end{equation}

In what follows, we will set $c_{0}$, $c_{1}$ = 1.

\begin{itemize}
\item{\bf{PREDICT} (without using Composer)}
    \begin{itemize}
        \item (1.3) On the axes provided to you, draw the state at time $t = 0$. As before, draw:
        \begin{itemize}
            \item The real part of the wavefunction.
            \item The imaginary part of the wavefunction.
            \item The probability density $|\psi_n(x)|^2$.
        \end{itemize}
        \item (1.4) Will this probability density evolve in time? Why?
    \end{itemize}
\item{\bf{EXPLORE} (using Composer)}
    \begin{itemize}
        \item Check your answer in Composer. Does this agree with what you predicted?
    \end{itemize}
\item{\bf{REVISE}}
    \begin{itemize}
        \item If your prediction was incorrect, can you use the visualizations from Composer that you explored in the previous question to re-think your understanding?
    \end{itemize}
\end{itemize}

\subsection*{The double-well system}

For this part of the exercise, we will need a new Composer file which will be opened by the interviewer. After the file has been opened, we will move beyond the single-well into the double-well system. This can be done by changing the value in the box labeled \textbf{Spacing between wells}. When this parameter is set to zero, there is only one well as shown currently in the Composer file. 

\subsubsection*{The states of the double-well system}

\begin{itemize}
\item{\bf{EXPLORE} (using Composer)}
    \begin{itemize}
        \item (2.1) Explore what happens to the energy levels that you see when you change \textbf{Spacing between wells} to other values in the range between 0 and 1.
        \item (2.2) How do the wavefunctions $\psi_n(x)$ for first two energy levels ($n = 0, 1$) change as you adjust the spacing between the wells?
    \end{itemize}
\end{itemize}

\subsubsection*{States localized to the left and right wells}

Now set \textbf{Spacing between wells} to 0.75.

Now we will explore superposition states. The goal is to add the $n = 0$ and $1$ states together to form a superposition state that exists only in one well or the other. A general superposition of states can be written as
\begin{equation}
    \psi_\mathrm{initial}(x) = c_0\psi_0(x) + c_1\psi_1(x).
\end{equation}

\begin{itemize}
\item{\bf{PREDICT} (without using Composer)}
    \begin{itemize}
        \item What are the values of $c_0$ and $c_1$ for a state:
        \begin{itemize}
            \item That exists only in the left well?
            \item That exists only in the right well?
        \end{itemize}
        \end{itemize}
\item{\bf{EXPLORE} (using Composer)}
    \begin{itemize}
        \item FOR THE LEFT WELL: Check your answer in Composer. Does this agree with what you predicted?
        \item FOR THE RIGHT WELL: Check your answer in Composer. Does this agree with what you predicted?
    \end{itemize}
\item{\bf{REVISE}}
    \begin{itemize}
        \item If your predictions were incorrect, can you use the visualizations from Composer that you explored in the previous question to re-think your understanding?
    \end{itemize}
\end{itemize}

\subsubsection*{Time evolution in the double-well system}

Now let's explore how time evolution happens in the double-well system.

\begin{itemize}
\item{\bf{PREDICT} (without using Composer)}
    \begin{itemize}
    \item (2.4) If we initialize the system in the left well, how will the state evolve in time? Why?
    \item (2.4) Can you predict how the time evolution changes as you change the separation?
    \end{itemize}
\item{\bf{EXPLORE} (using Composer)}
    \begin{itemize}
        \item Check your answer in Composer. Does this agree with what you predicted?
    \end{itemize}
\item{\bf{REVISE}}
    \begin{itemize}
    \item If your prediction was incorrect, can you use the visualizations from Composer that you explored in the previous question to re-think your understanding?
    \end{itemize}
\item{\bf{EXPLORE} (using Composer)}
    \begin{itemize}
    \item Explore how the time evolution changes as you change the well separation.
    \begin{itemize}
        \item (2.4) How do the dynamics change as you change the well separation?
        \item (2.5) What happens if the wells are very far apart?
    \end{itemize}
    \end{itemize}
\end{itemize}

\subsection*{Conclusion}

Thank you for your time here! Please leave your exercise sheet with us.
We very much appreciate your help!

\subsection{Full student trajectories}

\begin{figure}
  \includegraphics[page = 1, width=0.85\textwidth]{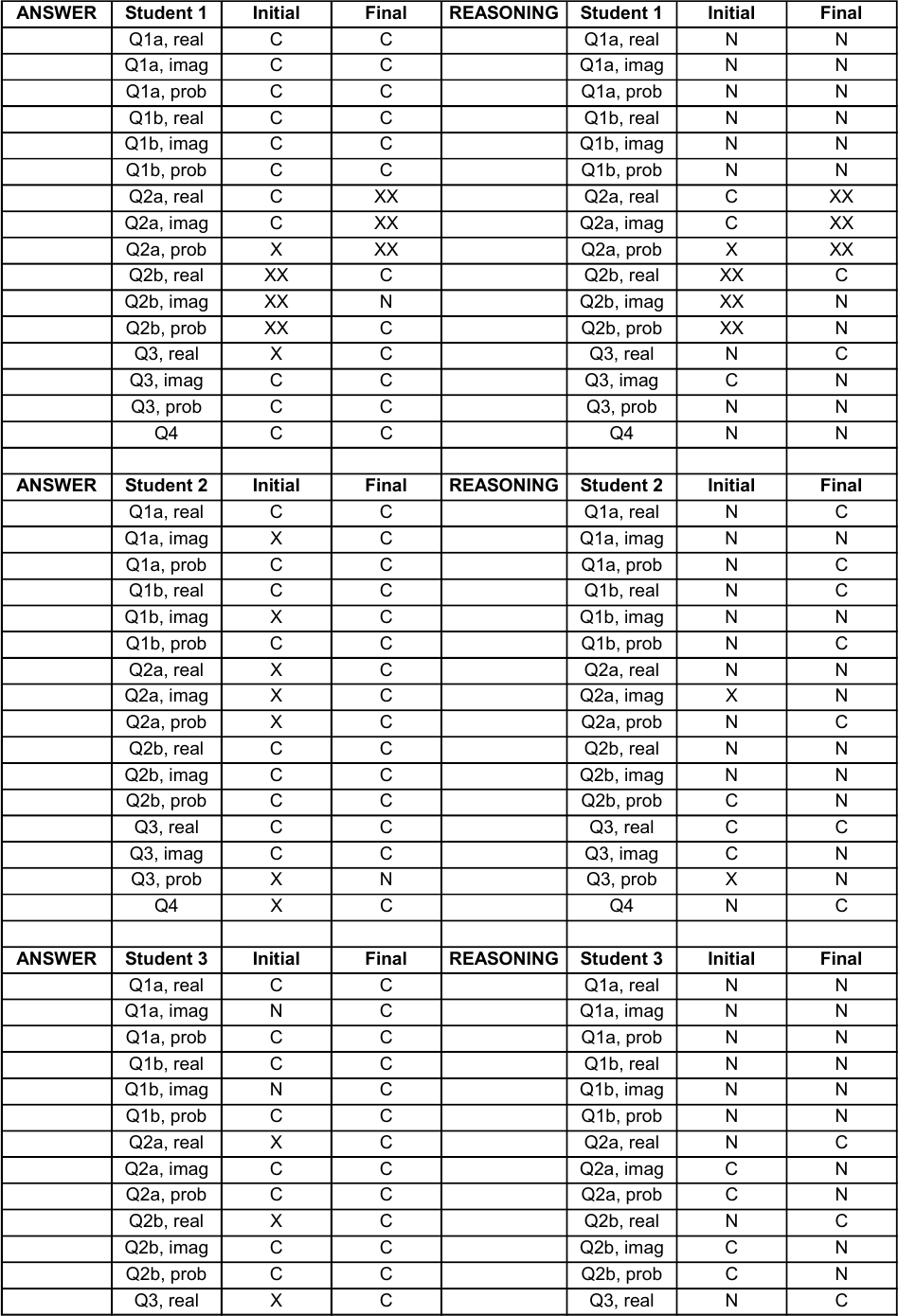}
  \caption{Full student trajectories, page 1. Student 3 continues on page 2.}
  \label{fig:trajectories1}
\end{figure}

\begin{figure}
  \includegraphics[page = 2, width=0.85\textwidth]{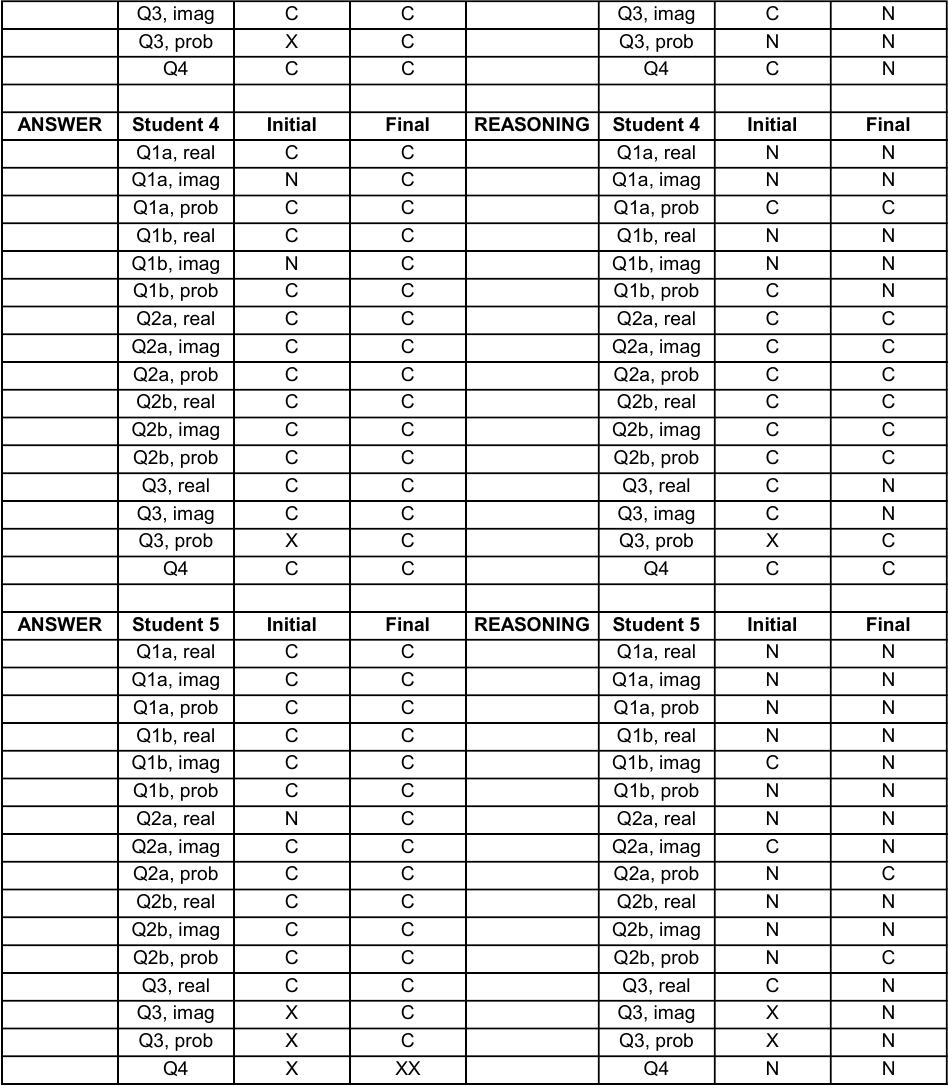}
  \caption{Full student trajectories, page 2.}
  \label{fig:trajectories2}
\end{figure}

In Figs.~\ref{fig:trajectories1} and \ref{fig:trajectories2}, we represent full student trajectories for the initial \emph{Predict} and final \emph{Explore/Revise} phases of the exercise. We track here both student answers and reasoning. In the figures, C corresponds to correct answer/reasoning, X to incorrect answer/reasoning, N to no answer/reasoning provided, and XX to a case where the student was using Composer incorrectly. Referring to the colors in Fig. 3 in the main text and reading the letters from left-to-right, yellow corresponds to CCNN; green to CCNC, XCNC, or XCXC; orange to CCCN or CCCC; red to XCXN, NCNN, XCNN or XNXN; and blue to anything with an XX in it.

%\newpage

\end{document}